\documentclass[aps,prl,twocolumn]{revtex4}
\usepackage{graphicx}
\usepackage{epstopdf}
\usepackage{epstopdf}

\begin{document}

\title{Influence of domain walls in the incommensurate charge density wave state of Cu intercalated 1$T$-TiSe$_2$}

\author{Shichao Yan$^{1}$, Davide Iaia$^{1}$, Emilia Morosan$^2$, Eduardo Fradkin$^1$, Peter Abbamonte$^1$ and Vidya Madhavan$^{1,*}$}

\affiliation{$^1$Department of Physics and Frederick Seitz Materials Research Laboratory, University of Illinois Urbana-Champaign, Urbana, Illinois 61801, USA}
\affiliation{$^2$Department of Physics and Astronomy, Rice University, Houston, Texas 77005, USA}

\begin{abstract}
We report a low-temperature scanning tunneling microscopy study of the charge density wave (CDW) order in 1$T$-TiSe$_2$ and Cu$_{0.08}$TiSe$_2$. In pristine 1$T$-TiSe$_2$ we observe a long-range coherent commensurate CDW (C-CDW) order. In contrast, Cu$_{0.08}$TiSe$_{2}$ displays an incommensurate CDW (I-CDW) phase with localized C-CDW domains separated by domain walls. Density of states measurements indicate that the domain walls host an extra population of fermions near the Fermi level which may play a role in the emergence of superconductivity in this system. Fourier transform scanning tunneling spectroscopy studies suggest that the dominant mechanism for CDW formation in the I-CDW phase may be electron-phonon coupling.
\end{abstract}

\maketitle Charge density wave (CDW) and superconductivity are two fundamental collective quantum states in solids. The interplay between these states and the nature of coexisting, competing phases in general are long standing questions in solid-state physics \cite{Net2001,Kiv1998,Cha2012,Soo1980}. 1$T$-TiSe$_2$ exhibits both CDW order and superconductivity which can be tuned by various parameters \cite{Kus2009,Li2016,Mor2006} making it an ideal system to study this interplay. At 202 K and ambient pressure, 1$T$-TiSe$_2$ undergoes a phase transition to a 2$\times$2$\times$2 commensurate CDW (C-CDW) order \cite{Sal1976} whose origin has been the subject of a long time debate  \cite{Ros2011,Mon2011,Mon2009,Cer2007, Hil2016}. Superconductivity emerges when the C-CDW phase is suppressed by applying pressure \cite{Kus2009}, electrostatic gating \cite{Li2016} or through Cu intercalation \cite{Mor2006}. Upon Cu intercalation for example, the C-CDW transition temperature quickly drops and the superconducting phase emerges from x $\sim$0.04 and reaches the maximal superconducting transition temperature of $\sim$4.2 K at x $\sim$0.08 \cite{Mor2006}. At first glance, this phenomenology suggests that CDW order and superconductivity are competing phases in this system \cite{Mor2006}. Recent studies however indicate that there might be a more exotic and complex interplay between them: X-ray diffraction (XRD) and electronic transport experiments report the emergence of an incommensurate CDW (I-CDW) phase which may play an important role in the emergence of superconductivity \cite{Li2016,Joe2014,Kog2016}.

Incommensuration may occur through two mechanisms: through a slight change of the CDW wavevector away from commensuration, or through the emergence of domains \cite{Bur1991,Tho1994,Mcm1975,Wu1989, Ma2016}. The idea that the I-CDW state in 1$T$-TiSe$_2$ occurs through the development of domains was first suggested by Y. I. Joe $et~al.$, based on XRD studies under high pressure. The authors further proposed that superconductivity first nucleates in the domain wall (DW) regions \cite{Joe2014}. A similar picture was used to explain the Little-Parks effect in the superconducting state of electrostatically gated 1$T$-TiSe$_2$ \cite{Li2016}. Very recently, based on XRD data, A. Kogar $et~al.$ reported an I-CDW phase near the superconducting dome in Cu$_x$TiSe$_2$ \cite{Kog2016}. These observations taken together strongly suggest that the I-CDW phase may be an important precursor to superconductivity in the 1$T$-TiSe$_2$ materials class. It is therefore critical to not only confirm the existence of DWs in the I-CDW phase of 1$T$-TiSe$_2$ but also to measure their effect on the local electronic structure. To do this, we use low-temperature scanning tunneling microscopy (STM) and spectroscopy (STS) to study pristine 1$T$-TiSe$_2$ and optimally doped Cu$_x$TiSe$_2$ (Cu$_{0.08}$TiSe$_2$). Note that all data were obtained at 6 K in the normal state.

\begin{figure}[]
\includegraphics[width=\columnwidth]{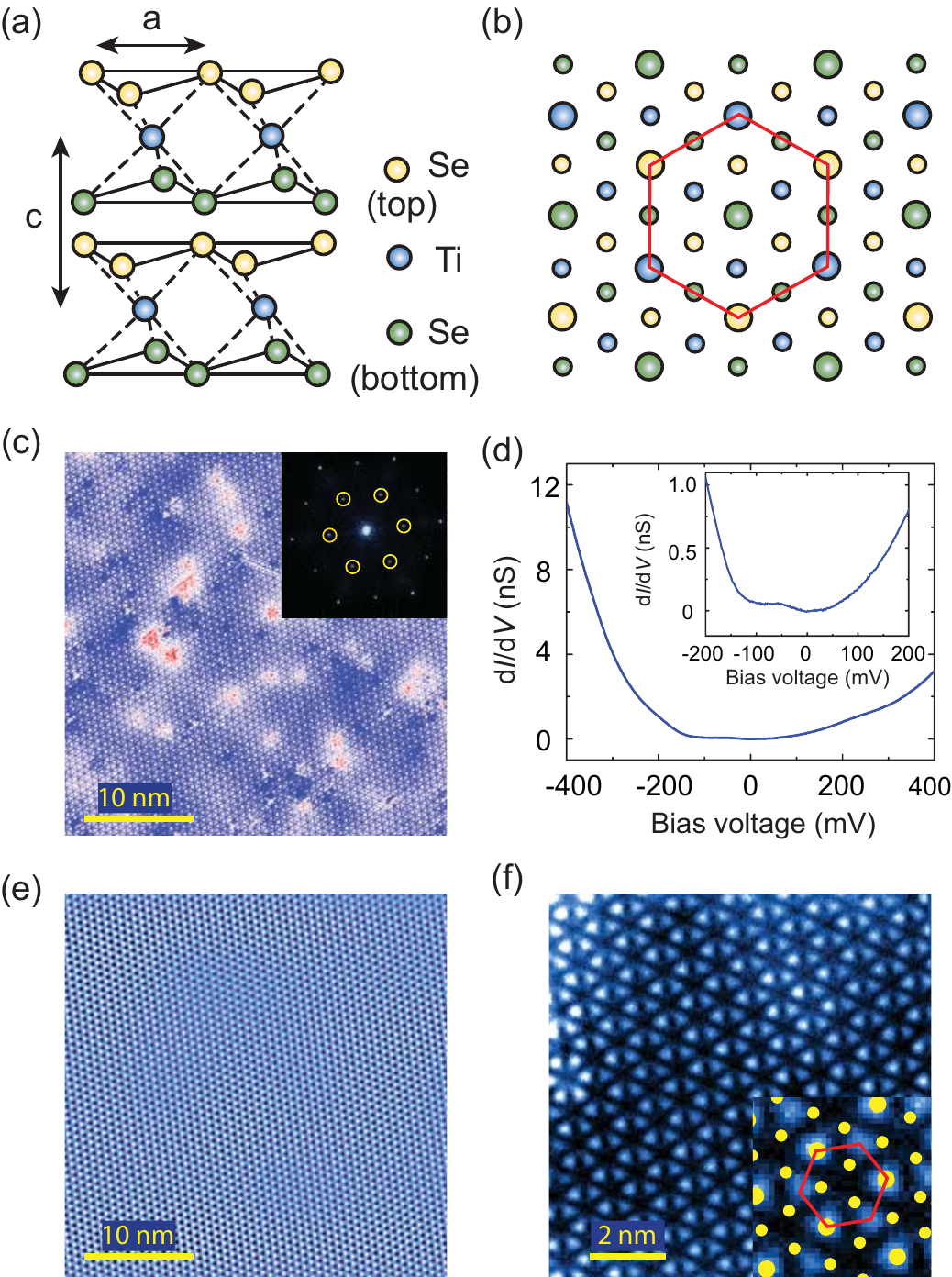}
\caption{(a) Crystal structure of the 1$T$-TiSe$_2$. (b) Schematic of the CDW order distribution in the Se-Ti-Se sandwich layer. The larger (smaller) circles represent the CDW maxima (minima). The red hexagon highlights the pattern formed by the CDW maxima in the top two (Se- and Ti-) layers. (c) STM constant current topography with $V_{\mathrm{s}} = -250{\,}\mathrm{mV}$, $I = 100{\,}\mathrm{pA}$. The inset shows its FT. The yellow circles indicate the CDW peaks. (d) Typical d$I$/d$V$ spectrum on TiSe$_2$. The inset shows the d$I$/d$V$ near $E_{\mathrm{F}}$. (e) Selective I-FT of CDW peaks shown in (c). (f) d$I$/d$V$ map over a 10 nm $\times$ 10 nm area at $V_{\mathrm{s}} = 100{\,}\mathrm{mV}$. The inset is a zoom-in d$I$/d$V$ map. The large (small) yellow dots represent the CDW maxima (minima) in the top Se layer. The red hexagon highlights the six-lobed hexagon. Set-up condition  is $V_{\mathrm{s}} = -400{\,}\mathrm{mV}$, $I = 1{\,}\mathrm{nA}$.}
\label{fig:Figure1}
\end{figure}

 1$T$-TiSe$_2$ consists of two-dimensional Se-Ti-Se sandwich layers in which the Se sheets have a hexagonal close-packed structure and the Ti atoms are in the octahedral centers defined by the two Se sheets (Fig.~1(a)). TiSe$_2$ cleaves between two such sandwich layers terminating in a Se surface. STM topography shows the top Se atoms, the surface 2$\times$2 superstructure corresponding to the CDW (seen as additional peaks at half of the Bragg reciprocal lattice vectors in the Fourier transform (FT, inset of Fig.~1(c))), and various native impurities seen as bright extended objects with trigangular/hexagonal symmetry (Fig.~1(c)) \cite{Hil2014}. Typically, in a CDW phase, one observes a gap in the density of states (DOS) near the Fermi level ($E_{\mathrm{F}}$). Fig.~1(d) shows a typical differential conductance (d$I$/d$V$) spectrum obtained on TiSe$_2$ away from the native atomic defects. From the change in slope around $-$110 mV and +10 mV, we deduce a partial gap energy scale of $\sim$120 mV, consistent with angle-resolved photoemission spectroscopy (ARPES) measurements of the band gap \cite{Kid2002,Zha2007,Qia2007}. While the DOS is certainly suppressed near $E_{\mathrm{F}}$, it remains finite and non-zero inside the gap, potentially due to impurity induced in-gap states.

To expose the charge distribution in the CDW phase and to separate it from the atomic corrugation it is necessary to look at d$I$/d$V$ map. From the d$I$/d$V$ map in Fig.~1(f) we find that the charge distribution displays a hexagonal structures with six lobes. The inset of Fig.~1(f) shows the expected CDW pattern corresponding to the top Se-layer (large and small yellow dots) superimposed on the six-lobed hexagon. Interestingly, while three of the hexagonal lobes are directly located on the CDW maxima of the top Se-layer, the other three lobes are located between three CDW minima (smaller yellow dots) of the top Se-layer. Comparing this to the schematic CDW pattern in Fig.~1(b), we conclude that our d$I$/d$V$ map reveals the CDW order in the top Se-layer as well as Ti-layer underneath.

STM images provide information on the ordering length scale and homogeneity of the CDW order. Examining the CDW pattern shown in Fig.~1(c) by eye, we conclude that it consists of a single domain. However, a better way to isolate the spatial characteristics of the CDW is to obtain a selective inverse Fourier transform (I-FT) of the CDW peaks in the FT. The resulting image (Fig.~1(e)) clearly shows a uniform CDW order over the 40 nm length scale of the image. In fact, I-FTs of areas as large as 115 nm (see Supplemental Materials (SM), \cite{SM}) show an equally uniform CDW pattern with no DWs indicating that despite the presence of intrinsic defects, the CDW phase is long-range ordered in this system.

We now investigate the fate of the CDW in Cu$_{0.08}$TiSe$_{2}$. Initial transport studies indicate that the C-CDW order parameter is heavily suppressed and is eventually destroyed as superconductivity emerges \cite{Mor2006}. However, an I-CDW phase coexisting with superconductivity has been proposed \cite{Li2016,Joe2014,Kog2016}. From the STM images (Fig.~2(a)) on Cu$_{0.08}$TiSe$_{2}$ we find a large number of atomic scale protrusions, which can be identified as Cu atoms or clusters on the surface \cite{SM}. The Cu atoms in the layer beneath can also be imaged at higher bias voltages and the observed density obtained from the layer underneath is consistent with a nominal doping of 8\% \cite{SM}.  From Fig.~2(a), we see that a CDW order persists in Cu$_{0.08}$TiSe$_{2}$  which at first glance looks very similar to 2$\times$2 CDW observed in the pristine samples. The FT image (inset, Fig.~2(a)) is however different from that of the parent compound. Instead of one peak each at the CDW wavevectors, we have a pair of CDW peaks in each direction. Taking the I-FT of these pairs we find that the resultant CDW pattern is extremely inhomogeneous (Fig.~2(b)). Tracking the CDW pattern across the inhomogeneous regions of the I-FT reveals that it may be due to phase shifts in the CDW pattern. This provides the impetus to carefully study high-resolution STM images (Fig.~2(c)) where we can now identify many DWs \cite{SM}. The DWs form long stripes (orange lines in Fig.~2(c)) and exist in all three equivalent directions in the sample. Zooming in to a single DW, we can see the $\pi$-phase shift across it \cite{SM}. Overall, this behavior is similar to the I-CDW phase observed in 1$T$-TaS$_2$ \cite{Bur1991,Tho1994,Wu1989,Nak1977}. Our data indicate that Cu-intercalation has changed the nature of the CDW from a commensurate to an incommensurate phase characterized by domains where the Cu atoms act as pinning impurities for the CDW. We note here that an I-CDW state with domains may be created either by a simple phase shift of the C-CDW between domains or by the appearance of two rotated I-CDW vectors which combine to produce domains \cite{SM}. From an analysis of the CDW structure inside the domains we are in the former case.

\begin{figure}[h]
\includegraphics[width=\columnwidth]{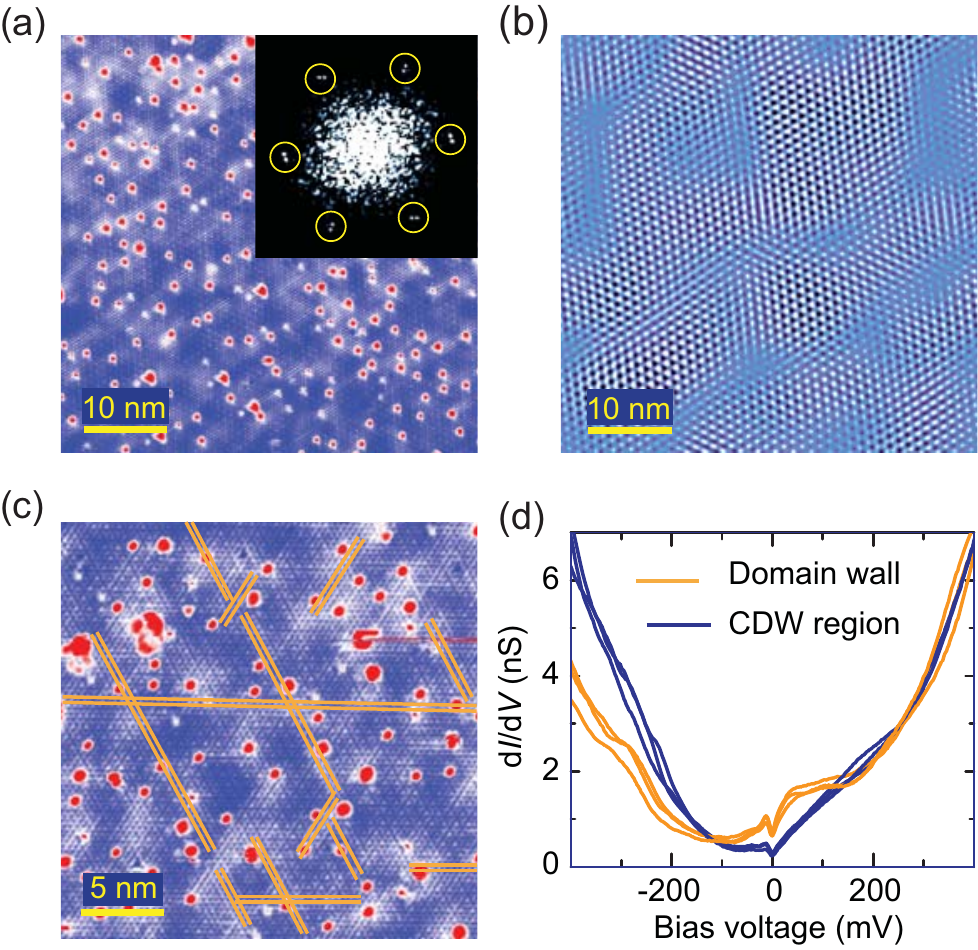}
\caption{(a) STM topography on Cu$_{0.08}$TiSe$_{2}$ with $V{\mathrm{s}} = -100{\,}\mathrm{mV}$, $I = 20{\,}\mathrm{pA}$. The inset is the FT of (a) and the yellow circles indicate the position of the CDW peaks. (b) I-FT by filtering the CDW components in the yellow circles of (a) inset. (c) STM topography with $V_{\mathrm{s}} = -150{\,}\mathrm{mV}$, $I = 20{\,}\mathrm{pA}$. The orange solid lines indicate the positions of the DWs. (d) d$I$/d$V$ spectra taken on randomly selected CDW regions and DWs. Set-up condition: 500 mV, 2 nA.}
\label{fig:Figure2}
\end{figure}

\begin{figure}[]
\includegraphics[width=\columnwidth]{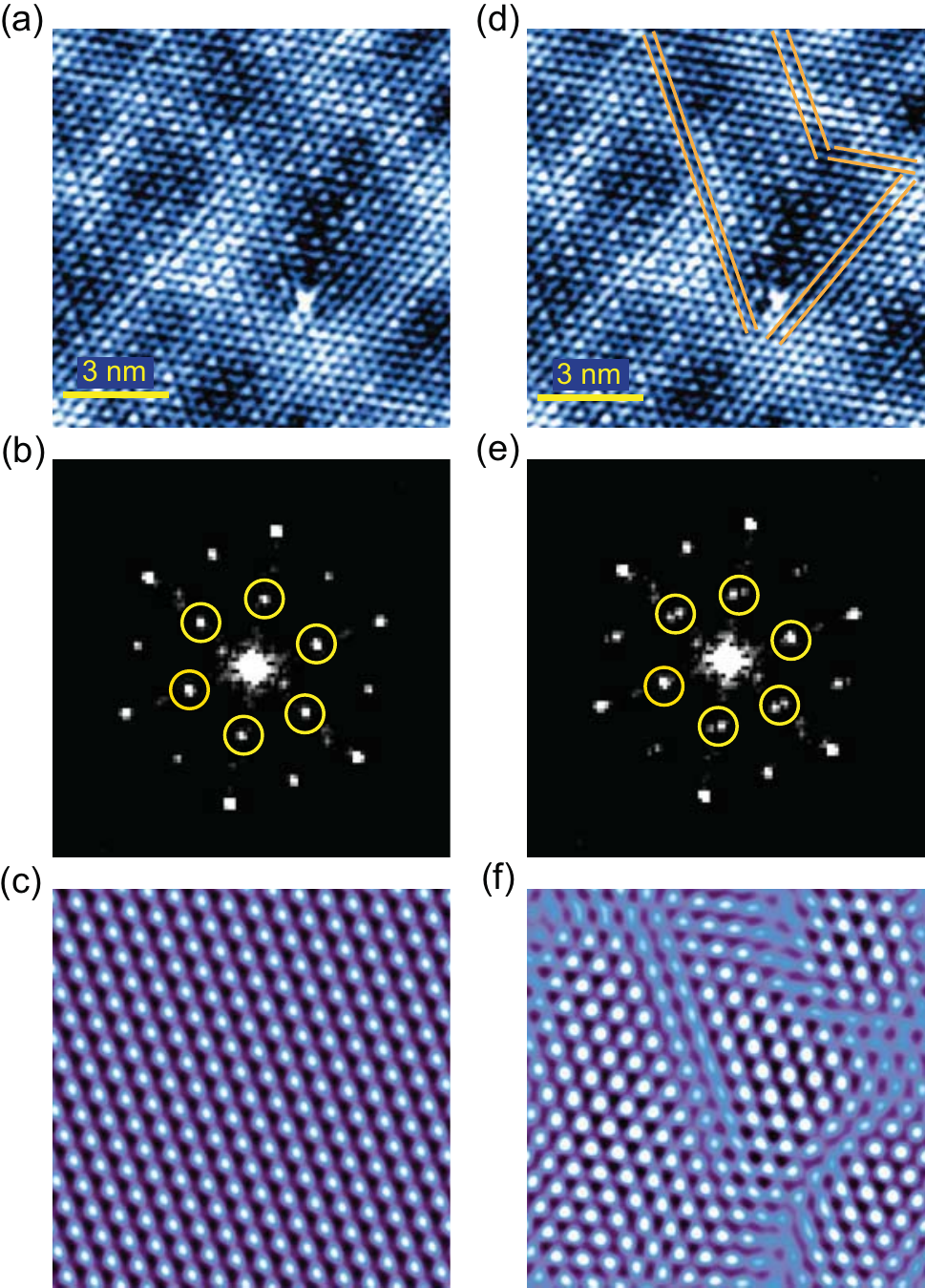}
\caption{(a) STM topography obtained on Cu$_{0.08}$TiSe$_{2}$ after removing the Cu atoms with STM tip. Set-up
condition: $V_{\mathrm{s}} = -150{\,}\mathrm{mV}$, $I = 20{\,}\mathrm{pA}$. (b) FT of (a). (c) The I-FT by filtering the CDW components in the yellow circles in (c). (d)-(f) The same as (a)-(c), but performed on the same area after DWs are created.}
\label{fig:Figure3}
\end{figure}

\begin{figure*}[]
\includegraphics[width=\textwidth]{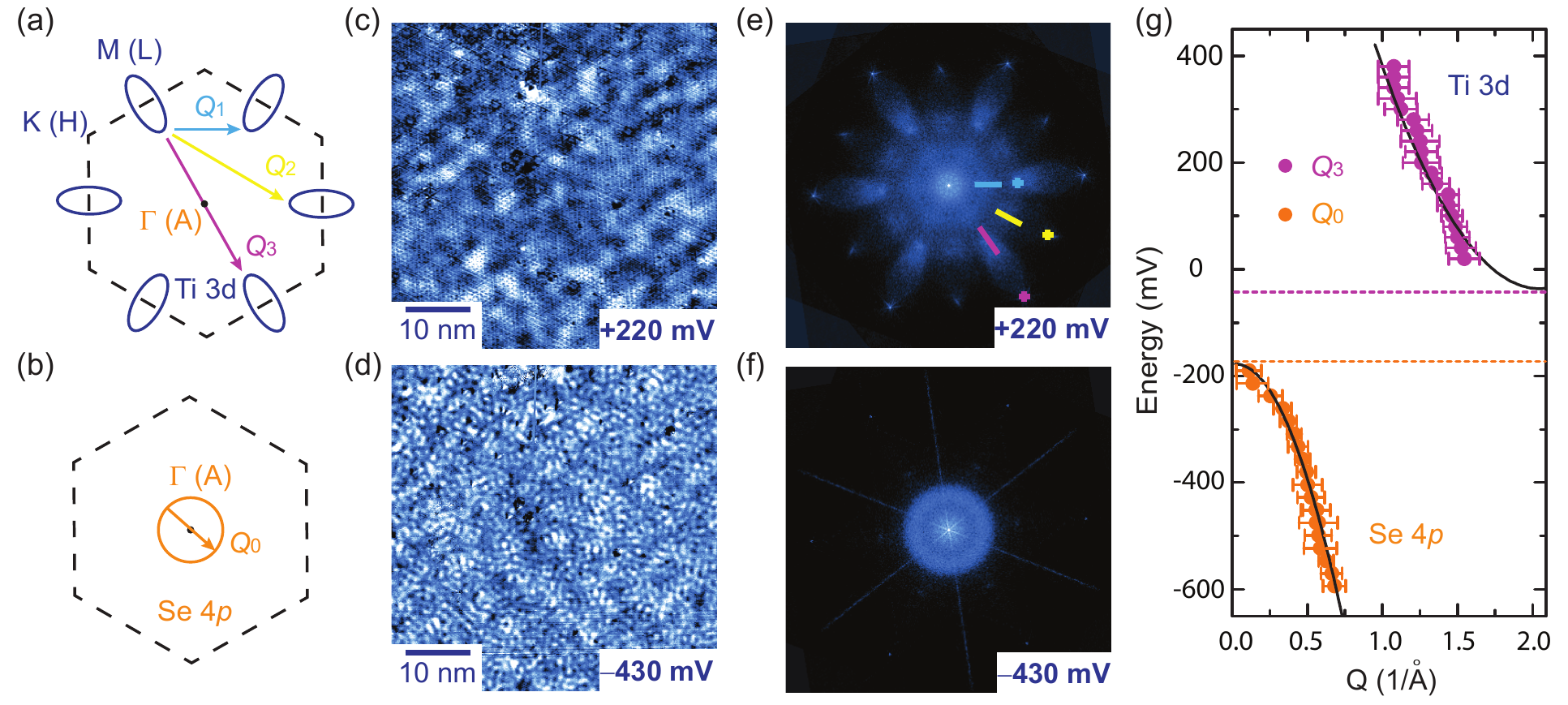}
\caption{(a) and (b)  Fermi surface topology of Ti-3d band (blue ellipses) and Se-4p band (orange circle). The dashed line is the first BZ. The arrows indicate the dominant scattering wavevectors. (c) and (d) Spatially resolved d$I$/d$V$ maps taken on Cu$_{0.08}$TiSe$_{2}$ sample at +220 mV and $-$430 mV. (e) and (f) Drift-corrected and symmetrized FTs of d$I$/d$V$ maps in (c) and (d). (g) Dispersions of $Q$$_{0}$ (orange dots) and $Q$$_{3}$ (pink dots) scattering vectors shown in (a) and (b).}
\label{fig:Figure4}
\end{figure*}

The discussion above suggests that domain structure in the I-CDW phase is accompanied by a splitting of CDW peaks in the FTs and obtaining I-FT images of the CDW peaks can be used to visualize this. To clarify this relationship, we performed a control experiment on the  Cu$_{0.08}$TiSe$_{2}$ surface. By moving the STM tip closer to the surface we find that we are able to remove Cu atoms \cite{SM}. Furthermore, we find that the domains in the Cu intercalated system can be perturbed by high STM bias voltages. Using these two techniques we now show that the peak splitting is a direct consequence of the presence of domains. Fig.~3(a) is an area of the Cu intercalated surface where the Cu atoms have been removed by the tip. This area shows a single domain, and the FT (Fig.~3(b)) shows a single set of CDW peaks. Correspondingly, the I-FT image of the CDW peaks shows a homogeneous CDW order (Fig.~3(c)). By scanning multiple times at a bias voltage of $-$350mV, we were able to create DWs (Fig.~3(d)) \cite{SM}. The FT of this perturbed image (Fig.~3(e)) shows that two of the three CDW peaks are now split. The third CDW peak remains un-split since there is only a very short section of a DW in this direction. The newly created CDW domains can also be clearly resolved in the I-FT image (Fig.~3(f)). This cements relationship between the split peaks and the domains and confirms that obtaining selective I-FT images is a good tool to capture the spatial structure of the DWs. A simple mathematical description of peak splitting due to domains can be found in SM. The magnitude of splitting provides an average length scale of $\sim$10 nm for the in-plane domain size \cite{SM}, a length scale similar to XRD measurements of $\sim$13 nm for $c$-axis domains \cite{Kog2016}.

Given the proposal that superconductivity might nucleate in the DWs \cite{Li2016,Joe2014,Kog2016}, the natural question is: what is the effect of DWs on the local DOS? Unlike impurities that perturb the lattice and electronic structure by adding potentials or strains, the DWs seen by us represent topological defects in the arrangement of charge. Any effect of such DWs on the electronic structure is therefore expected to have a non-trivial origin.  As shown in Fig.~2(d), the spectra on DWs show an enhanced DOS near  $E_{\mathrm{F}}$  compared to spectra within the localized CDW regions. This can also be seen in d$I$/d$V$ maps at low energies where DWs appear as high intensity lines \cite{SM}. This intriguing observation indicates that the DWs host an extra population of fermions. Moreover, in the particular case of a period two CDW, the CDW order parameter is expected to go to zero at the DWs. The higher DOS combined with a suppression of the CDW at DWs may be the key factors that aid the emergence of superconductivity in this system. 

Next we explore possible mechanisms for CDW formation in Cu$_{0.08}$TiSe$_{2}$ by using FT of STM d$I$/d$V$ maps (FT-STS) to extract the band structure \cite{SM, Hof2002}. FT-STS at a given energy contains the allowed scattering vectors ($Q$-vectors) between the k-space electronic states within the constant energy contour (CEC) at that energy. We obtain the energy-dispersion relation by tracking $Q$-vectors magnitudes with energy. Note that due to the propensity of surface Cu atoms to be moved by the tip, it is not possible to obtain noise free d$I$/d$V$ maps with Cu atoms present. d$I$/d$V$ maps were therefore obtained on areas where Cu atoms were deliberately removed by the tip (Figs.~4(c) and 4(d)). Spectra taken on the cleaned surfaces are almost identical to those on Cu covered surfaces \cite{SM} and the CDW remains incommensurate. This indicates that much of the band structure is bulk-like, determined by the doping in the bulk of the sample.

At energies near $E_{\mathrm{F}}$ the band structure is dominated by two bands: a Ti-3d derived band around the L point (at the Brillouin zone (BZ) edge) and an Se-4p derived hole-like band close to the $\Gamma$ point (the center of the BZ), (Figs.~4(a) and 4(b)). The Se-4p bands have circle-like CECs, and the $Q$-vectors corresponding to intra pocket scattering form a ring in the FT-STS ($Q$$_{0}$, Figs.~4(b) and 4(f)). Upon changing the sample voltage from $-$600 mV toward $E_{\mathrm{F}}$, the ring-like feature in the FT-STS gets continuously smaller and vanishes around $-$180 mV. No clear dispersive pattern is observed between $-$180 mV and $E_{\mathrm{F}}$. For the Ti-3d band, the CECs consists of six elliptical electron pockets, and there are three main sets of scattering wave vectors ($Q$$_{1}$, $Q$$_{2}$, $Q$$_{3}$) which represent the scattering between elliptical pockets along $\Gamma$M, $\Gamma$K and $\Gamma$M, respectively (Figs.~4(a) and 4(e)). As energy is increased from $E_{\mathrm{F}}$ to +400 mV, the sizes of the six elliptical pockets increase and the resulting scattering vectors, $Q$$_{1}$, $Q$$_{2}$, $Q$$_{3}$ move towards the BZ center \cite{SM}.

 We focus here on the two strong scattering vectors, $Q$$_{0}$ and $Q$$_{3}$, and their dispersions (Fig.~4(g)). The positions for the top of the Se-4p band and the bottom of the Ti-3d band are extrapolated by parabolic fit to the dispersion. The difference between the valence band top at $\sim$$-$170 meV and conduction band bottom band at $\sim$$-$40 meV gives us a band gap of $\sim$130 meV, consistent with the previous ARPES measurements \cite{Zha2007,Qia2007}. Our measurements allow us for the first time to directly correlate the I-CDW state with the band structure. Our data indicate that Cu-intercalation moves the Fermi energy deeper into the conduction band compared to the pristine samples, thereby removing the nesting condition at $E_{\mathrm{F}}$.  This rules out Fermi surface nesting as the mechanism for the observed I-CDW. Many studies have suggested that there are both excitonic and phononic contributions to the C-CDW in pristine 1$T$-TiSe$_2$ \cite{Wez2010,Wez2010a,Por2014}. While the electron doping into the Ti-3d band suggests that the excitonic contribution should be weakened, the electron-phonon coupling should be less affected. Our data therefore indicate that in contrast to the pristine case where excitonic and phononic contributions are both implicated, in Cu$_{0.08}$TiSe$_{2}$ electron-phonon interactions may play a dominant role in the formation of the I-CDW \cite{Kog2016}. 

In conclusion, our data clearly show that the incommensuration due to Cu intercalation proceeds through DW formation. The emergence of the I-CDW phase as observed by us can be used to explain the loss of long-range coherence of C-CDW phase above the superconducting dome observed in ARPES measurements \cite{Zha2007,Qia2007}. We conclude that the I-CDW phase and associated DWs should be a common element of 1$T$-TiSe$_2$ samples that exhibit superconductivity through doping, gating, or pressure. The enhancement of DOS at the domain walls may be a crucial element in the emergence of superconductivity. Further STM studies of these samples below the superconducting transition temperature would be important in fully understanding the role of the I-CDW phase in superconductivity.

$^*$ vm1@illinois.edu

\end{document}